\documentclass[]{JHEP3}
\usepackage{graphicx}
\JHEPspecialurl{http://jhep.sissa.it/JOURNAL/JHEP3.tar.gz}

\usepackage{amsmath}
\usepackage{epsfig,multicol,bbm}
\usepackage{url}

\newcommand{\newc}{\newcommand}
\newc{\gsim}{\lower.7ex\hbox{$\;\stackrel{\textstyle>}{\sim}\;$}}
\newc{\lsim}{\lower.7ex\hbox{$\;\stackrel{\textstyle<}{\sim}\;$}}
\newc{\gev}{\,{\rm GeV}}
\newc{\mev}{\,{\rm MeV}}
\newc{\ev}{\,{\rm eV}}
\newc{\kev}{\,{\rm keV}}
\newc{\tev}{\,{\rm TeV}}

\newc{\mz}{M_Z}
\newc{\mpl}{M_*}
\newc{\mw}{m_{\rm weak}}
\newc{\nr}[1]{N^c_R{}_{#1}}
\usepackage{amsmath}
\def\beq{\begin{equation}}
\def\eeq{\end{equation}}
\def\bea{\begin{eqnarray}}
\def\eea{\end{eqnarray}}
\newcommand{\beqa}{\begin{eqnarray}}
\newcommand{\eeqa}{\end{eqnarray}}
\def\bitem{\begin{itemize}}
\def\eitem{\end{itemize}}
\newc{\ie}{{\it i.e.}}          \newc{\etal}{{\it et al.}}
\newc{\eg}{{\it e.g.}}          \newc{\etc}{{\it etc.}}
\newc{\cf}{{\it c.f.}}

\def\bar#1{\overline{#1}}

\def\inv{^{\raise.15ex\hbox{${\scriptscriptstyle -}$}\kern-.05em 1}}
\def\lbar{{\lower.35ex\hbox{$\mathchar'26$}\mkern-10mu\lambda}} 

\def\to{\rightarrow}

\let\<=\langle
\let\>=\rangle

\let\+=\uparrow

\newcommand\fverb{\setbox\fverbbox=\hbox\bgroup\verb}
\newcommand\fverbdo{\egroup\medskip\noindent%
			\fbox{\unhbox\fverbbox}\ }
\newcommand\fverbit{\egroup\item[\fbox{\unhbox\fverbbox}]}
\newbox\fverbbox

\title{
Split families unified}

\date{\today}

\author{Nathaniel Craig\\
Department of Physics and Astronomy, Rutgers University, Piscataway, NJ 08854\\
	School of Natural Sciences, Institute for Advanced Study, Princeton, NJ 08540\\
		E-mail: \email{ncraig@ias.edu}}
	
\author{Savas Dimopoulos\\
	Department of Physics, Stanford University, Stanford, CA 94305\\
	E-mail: \email{savas@stanford.edu}}
	
\author{Tony Gherghetta\\
       ARC Centre of Excellence for Particle Physics at the Terascale, 
	School of Physics, University of Melbourne, Victoria 3010, Australia\\
	Department of Physics, Stanford University, Stanford, CA 94305 \\
	E-mail: \email{tgher@unimelb.edu.au}}

\preprint{RUNHETC-2011-25, SU-ITP-11/52}	

\abstract{We present a simple supersymmetric model of split families consistent with flavor limits
that preserves the successful prediction of gauge coupling unification and naturally accounts for the 
Higgs mass. The model provides an intricate connection between the Standard Model flavor hierarchy, 
supersymmetric flavor problem, unification  and the Higgs mass. In particular unification favors a naturally 
large Higgs mass from D-term corrections to the quartic couplings in the Higgs potential. The unification 
scale is lowered with a stable proton that can account for the success of $b-\tau$ Yukawa coupling unification.
The sparticle spectrum is similar to that of natural supersymmetry, as motivated by the supersymmetric 
flavor problem and recent LHC bounds, with a heavy scalar particle spectrum except for a moderately 
light stop required for viable electroweak symmetry breaking. Finally, Higgs production and decays, NLSP 
decays, and new states associated with extending 
the Standard Model gauge group above the TeV scale provide signatures for experimental searches at the LHC.

}


\begin{document} 

\maketitle

\section{Introduction}

In every extension of the Standard Model (SM) addressing the hierarchy problem, there is a tension between naturalness and flavor. This led to the early demise of extended technicolor theories. In the supersymmetric Standard Model \cite{Dimopoulos:1981zb} this tension was addressed by postulating universality -- the equality of the masses of sparticles of the same gauge and spin quantum numbers at the Grand Unified Theory (GUT) scale. The Large Hadron Collider (LHC) has recently constrained the particle masses in these theories to lie roughly above a TeV. This again increases the tension between naturalness and observation, and raises the possibility that even the 14 TeV LHC may fail to discover supersymmetry, as can easily happen in theories of split supersymmetry \cite{ArkaniHamed:2004fb, Giudice:2004tc}.

Fortunately, there is another way to maintain naturalness and remain in agreement with experiment while allowing for the possibility of light sparticles observable at the LHC with increased integrated luminosity: the first two generations of sparticles are heavy and the third generation is light. The essential mechanism was proposed in the LEP era  \cite{Dimopoulos:1995mi,  Pomarol:1995xc, Cohen:1996vb} as an alternate way to align the absence of FCNCs with naturalness. It is based on the observation that naturalness is sensitive mostly to the top quark -- and more broadly, the third generation -- whereas the largest contribution to FCNCs involves the first two generations. Even so, in this scenario the first two generations have to be approximately degenerate and in the multi-TeV regime to satisfy constraints from FCNCs \cite{Barbieri:2011ci} while the third generation particles, such as the stop, can easily be near the top mass. Although these ideas were originally motivated by flavor, they have found new appeal in the LHC era due to the non-observation of first-generation squarks below a TeV.  Despite limits on scalar partners of the first two generations, the LHC constraint on the mass of the third generation squarks is much weaker; stop  masses as light as $\sim 300$ GeV remain feasible \cite{Kats:2011qh, Papucci:2011wy, Brust:2011tb, Essig:2011qg}.  In light of data, a natural supersymmetric solution to the hierarchy problem would seem to favor theories with light third generation scalars and heavier first- and second generation scalars. We will refer to these theories -- the focus of this work -- as theories of ``split families'' or ``natural supersymmetry''.

In addition hints of a  Higgs boson at 125 GeV \cite{Chatrchyan:2012tx,:2012si} provide an independent source of tension with naturalness in the MSSM; to accomplish this one needs either very heavy stops or large $A$-terms, and either possibility pushes against naturalness \cite{Draper:2011aa}. Thus in order to be compatible with natural
supersymmetry new contributions to the quartic couplings in the Higgs potential are required. The most common
way to extend the MSSM is to either introduce a new singlet which couples to the Higgs doublets~\cite{Hall:2011aa} 
or rely on non-decoupling $D$-terms~\cite{Batra:2003nj}. In either case, the successful prediction of gauge coupling unification in the MSSM is not automatic. Thus it represents a challenge to preserve gauge coupling unification while simultaneously increasing the Higgs mass and explaining the origin of natural supersymmetry.

The purpose of this paper is to propose a theory of natural supersymmetry consistent with flavor limits that preserves the prediction of gauge coupling unification -- the one quantitative success of physics beyond the SM -- and naturally accounts for the mass of the Higgs.  The theory provides a weakly coupled UV completion of \cite{Craig:2011yk}. In this theory flavor, sflavor, unification, and the Higgs mass are intricately inter-related phenomena. Above the TeV scale the Standard Model gauge group is split into two SM gauge groups connected by link fields. The third generation and the Higgs doublets are charged under one SM group, while the first two generations are charged under the second SM group.  Remarkably the extra states required to address the Standard Model flavor hierarchy cause the two SM groups to separately unify, in a form of accelerated unification~\cite{ArkaniHamed:2001vr}. Even though the
unification scale is generically lowered, the proton remains stable as long as the unification scale for the third generation remains above $\sim 10^{12}$ GeV. Proton decay involving the first two generations is prevented by 
relying on a deconstructed orbifold mechanism, whereby the first two generation fermions are not charged under the unified gauge group. This elegantly explains why Yukawa coupling unification occurs only for the third generation 
$(b-\tau)$ and not the first two generations. Alternatively the two SM gauge groups can unify at a common unification scale $\sim 10^{10}$ GeV. This is consistent with proton decay by also requiring that the third generation remains uncharged under the unifying gauge group. 

The requirement of successful gauge coupling unification favors large $D$-term corrections to the quartic couplings in the Higgs potential which increases the Higgs mass.  A naturally split sparticle spectrum occurs by transmitting supersymmetry breaking to the first two generations via gauge mediation, with the link fields providing a further suppression to the third generation. Interestingly, to achieve viable electroweak symmetry breaking the stops are only moderately light. Likewise, the link fields provide a natural origin for the Higgs-sector parameters $\mu$ and $B \mu$. 
In addition there are a number of phenomenological consequences, including D-term corrections 
for Higgs production and decays, NLSP decays with a gravitino LSP, and -- in contrast with many other theories of natural supersymmetry -- the possibility of observing new states at the LHC.

The outline of the paper is as follows. In Section~\ref{sec:decon} we present a simple model, based on deconstruction, that simultaneously addresses the issues of flavor and unification in a setup that naturally accommodates split families. The Higgs sector is discussed in Section 3, where the implications of large $D$-term corrections to the quartic couplings are addressed for the Higgs mass, Higgs production, and decays. In Section 4 we briefly discuss the phenomenology of the sparticle spectrum, including the possibility of observing new light states that distinguish our low-scale framework from models based on dynamics at a high scale. We conclude in Section 5 by comparing the virtues of various approaches to natural supersymmetry.

\section{A Simple Model} \label{sec:decon}

The underlying idea is that the Standard Model gauge group $G_{SM}$ is extended to two (or conceivably more) copies, higgsed down to the diagonal at an appropriate scale. Below the scale of higgsing, the theory is essentially that of the Standard Model; above the scale of higgsing, the Standard Model splits into a series of distinct gauge groups connected by bifundamental link fields. As the number of gauge groups is increased, such constructions reproduce the physics of an extra dimension, and indeed provide a well-behaved UV completion to various extra-dimensional theories. But there is no need to commit to a specific UV picture; in some sense the simple deconstructed model represents a general effective theory for many possible models that realize the minimal supersymmetric spectrum, whether four- or five-dimensional in origin.

Consider the simplest deconstruction, consisting of gauge groups $G_A \times G_B$ connected by bifundamental link fields $\chi, \tilde \chi$ and interactions that lead to vacuum expectation values $\langle \chi \rangle, \langle \tilde \chi \rangle \neq 0$  that higgs $G_A \times G_B \to G_{SM}$ at an appropriate scale.\footnote{The higgsing may occur due to supersymmetric~\cite{Cheng:2001an} or nonsupersymmetric~\cite{ArkaniHamed:2001vr} terms in the potential, the details of which are not crucial to our purposes. The supersymmetric case requires an additional adjoint chiral superfield charged under $G_B$, while the nonsupersymmetric case does not; in what follows we will assume the matter content of the nonsupersymmetric case. Both approaches require gauging an additional (spontaneously broken) $U(1)$ symmetry under which the link fields are charged in order to stabilize a flat direction. Due to flavor considerations the $U(1)$ must be extended to include some first- and second-generation Standard Model superfields, but this does not alter the low-energy phenomenology. For an alternative possibility involving trinification without an additional $U(1)$, see \cite{Maloney:2004rc}.} 
The most compelling scale of higgsing is low, around $1-20$ TeV.  In order to embed the Standard Model gauge fields in a diagonal subgroup of $G_A \times G_B$, we require $G_A, G_B \supset SU(3) \times SU(2) \times U(1)$.  Spontaneous symmetry breaking leads to one set of massless $SU(3)_C \times SU(2)_L \times U(1)_Y$ gauge bosons, as well as a set of heavy gauge bosons of mass $M_i^2 = 2 (g_{A_i}^2 + g_{B_i}^2) \langle \chi \rangle^2.$

We imagine that supersymmetry breaking is communicated primarily to fields charged under $G_B$. Perhaps the simplest and most attractive means of accomplishing this is via messengers of gauge mediation charged under $G_B$. As is typically the case with dynamical models of supersymmetry breaking, we assume there is an approximate $R$-symmetry that suppresses gaugino masses relative to the leading-order scalar soft masses.
Given this setup, fields charged under $G_B$ receive flavor-blind soft masses of order $
m_{GM}^2 \sim \left( \frac{\alpha}{4\pi} \right)^2 \left( \frac{F}{M}\right)^2 $
where $M$ is the messenger scale. In contrast, the two-loop mass-squared of scalars charged under $G_A$ are further suppressed by a schematic factor $\langle \chi \rangle^2 / M^2$. The additional suppression for fields charged under $G_A$ arises because their two-loop soft masses only arise {\it below} the scale $\langle \chi \rangle$.\footnote{In fact, important contributions arise at both two and three loops; as we will discuss below, the three-loop contributions dominate when $\langle \chi \rangle /M < 4 \pi$, as is typically the case here.} 

\begin{figure}[h]
   \centering
   \includegraphics[width=4in]{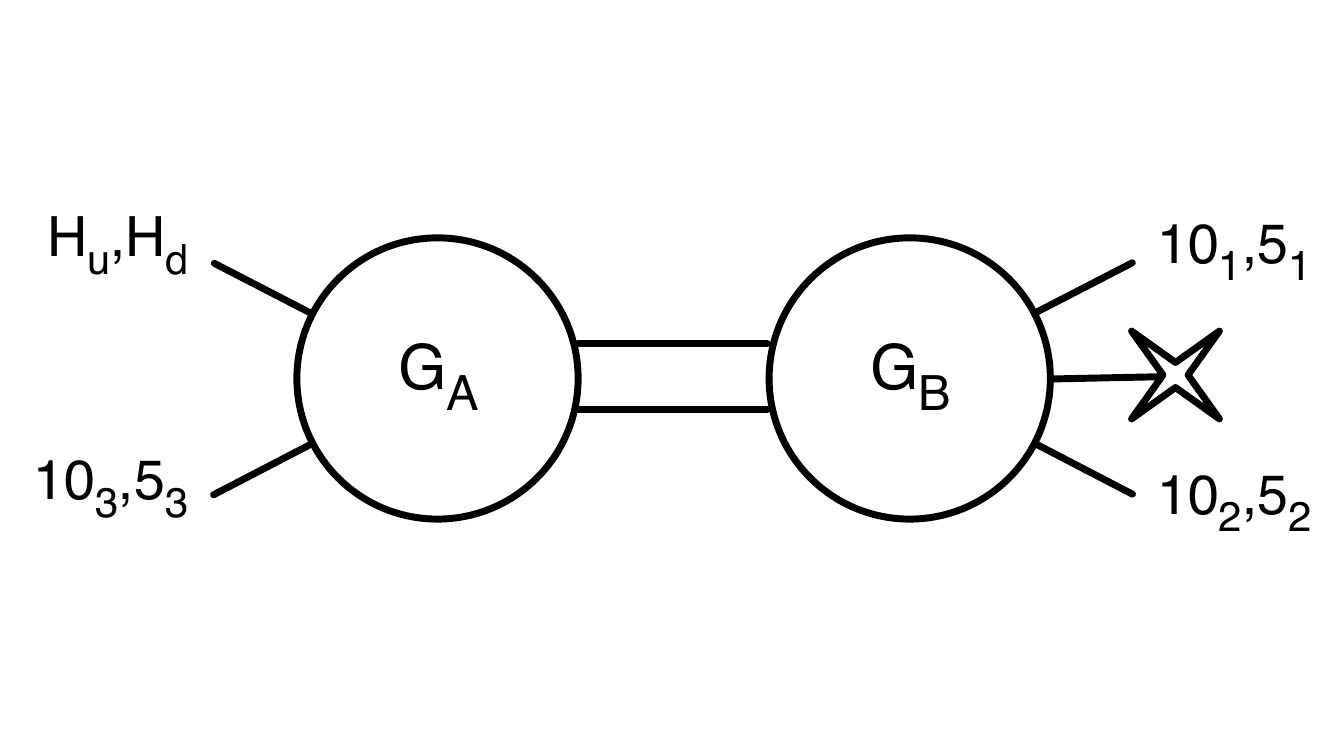}
   \caption{The deconstructed model. }
   \label{fig:decon}
\end{figure}

Motivated by the largeness of the top Yukawa coupling; the intimate connection between the top quark and the Higgs; and the constraints of naturalness, we charge the third generation chiral superfields and $H_u, H_d$ under $G_A$, while charging the first two generations under $G_B.$ The model is shown schematically in Fig.~\ref{fig:decon}. 

\subsection{Flavor}

Gauge-group locality determines the structure of both fermionic and sfermionic flavor. Since only the third generation superfields and the Higgs multiplets are charged under $G_A$, only the Yukawa interactions of the third generation are marginal operators. Yukawa couplings involving fields of the first two generations may arise via irrelevant operators upon insertions of the link field vevs, as discussed in detail in \cite{Craig:2011yk} (see also \cite{Auzzi:2011eu}). Such irrelevant operators arise from integrating out massive matter at the scale $M_* \sim M$. 

Crucially, the matter representations required by a complete theory of flavor have strong implications for the prospects of gauge coupling unification. First, consider the matter required to generate the Yukawa terms $(Y_d)_{31}$ and $(Y_d)_{32},$ vital for a realistic CKM matrix. These may be generated via marginal operators of the form
\beq
W \supset H_d Q_3  \bar d' + d' \bar d_2 \tilde \chi_h + M_* d' \bar d'  + \dots
\eeq
where $\bar d'$ has quantum numbers of a $\bar{d}$ matter superfield charged under $G_A$, and we use the notation of \cite{Craig:2011yk} for the representations of the link fields $\chi, \tilde \chi.$ Similar textures for the lepton Yukawas may be generated by integrating out the analogous field $L'$ with quantum numbers of an $L$ superfield. Together, $\bar d', L'$ and $d', \bar L'$ fill out a complete ${\bf 5 + \bar 5}$ multiplet under $SU(5)_A$. These fields have the quantum numbers of a unified vector-like fourth generation, and introduce no new contributions to proton decay at dimension 5 or lower.

In contrast, the Yukawa entries for the first two generations may be generated by operators of the form
\beq
W \supset H_u \tilde \chi_l H_d' + H_u' Q_2 \bar u_2+ M_*H_u' H_d' + \dots
\eeq
Here $H_u', H_d'$ have the quantum numbers of the doublets $H_u, H_d$ under $G_B$. In this case, it is important that there are no analogs of Higgs triplets, since such fields would introduce new (and prohibitively large) sources of dimension-five proton decay. The fields $d', \bar d', L', \bar L', H_u', H_d'$ are sufficient to generate realistic Yukawa textures.  Taken together, this suggests that a successful theory of flavor requires the addition of a vector-like pair of unified multiplets to $G_A$ and a vector-like pair of $SU(2)$ doublets to $G_B$. Thus flavor has strong implications for unification.

\subsection{Unification}

A striking feature of the supersymmetric extension of the Standard Model is that the extra states in the form of Higgsinos and gauginos cause the gauge couplings to unify at a high scale $M_{GUT}\sim 2\times 10^{16}$ GeV. This provides indirect 
evidence for supersymmetry and suggests that gauge coupling unification should also be preserved in natural supersymmetric extensions of the MSSM. 

In our setup the Standard Model gauge couplings, $g_i$ measured at some low-energy value $\mu_0$ (usually $m_Z$), 
are run up through a supersymmetric threshold to the scale of higgsing, $\langle \chi \rangle$. At this scale they are matched 
onto the set of gauge couplings $g_{A_i}$ and $g_{B_i}$ associated with the gauge groups $G_A$ and $G_B$, respectively 
via the relation 
\beq
\label{eqn:couplings}
    \alpha_i^{-1} (\langle \chi \rangle) =  \alpha_{A_i}^{-1}(\langle \chi \rangle) +\alpha_{B_i}^{-1} (\langle \chi \rangle)~,
\eeq
where $\alpha_I=g_I^2/(4\pi)$ for $I=i, A_i,B_i$  and $i=1,2,3$.
The running of all gauge couplings is simply determined by the one-loop renormalization equation
\beq
\label{rgeqn}
    \frac{d \alpha_{I}^{-1}}{d \log \mu} = -\frac{b_{I}}{2\pi}~,
\eeq
where $b_I$ are the $\beta$ function coefficients, and $\mu$ is the renormalization scale.

To understand unification in the two-site model let us first recall how unification occurs in the MSSM. Unification is 
characterized by two parameters $\alpha_{GUT}$ and $M_{GUT}$ which are then determined in terms of the low-energy values 
$\alpha_{1,2}(\mu_0)$ by using two of the three equations in (\ref{rgeqn}). This is simply the statement that two lines 
with different slopes will eventually cross. However once $\alpha_{GUT}$ and 
$M_{GUT}$ are fixed we can use the remaining equation to obtain a nontrivial prediction for $\alpha_3(\mu_0)$, namely
\beq
\label{a3inv}
     \alpha_{3}^{-1}(\mu_0) = (1+R) \alpha_{2}^{-1}(\mu_0) -R\alpha_{1}^{-1}(\mu_0)
     = \frac{12}{7} \alpha_{2}^{-1}(\mu_0) -\frac{5}{7}\alpha_{1}^{-1}(\mu_0)~,
\eeq
where 
\beq
\label{Bratio}
      R\equiv \frac{b_3-b_2}{b_2-b_1} = \frac{5}{7}\simeq 0.714~.
\eeq
In the MSSM the prediction (\ref{a3inv}) remarkably agrees with the measured value at $\mu_0=M_Z$.

In the two-site model it may seem that with separate $A$ and $B$ couplings we have lost this remarkable MSSM 
prediction. However we will see that the couplings still unify and a nontrivial prediction can still be obtained.
We begin with the assumption that unification occurs on the $A$ site. Using two of the three $A$ equations we can 
determine the unified coupling, $\alpha_A$ and the unification scale $M_A$ parameters in terms
of the arbitrary initial values of $\alpha_{A_{1,2}}$ at the higgsing scale. Via the relation (\ref{eqn:couplings}) the $B$ 
initial values $\alpha_{B_{1,2}}$ are then fixed. Since the $B$ couplings run separately from the $A$ couplings we can 
again trade the initial $B$ values $\alpha_{B_{1,2}}$ at the higgsing scale for the value of the unified coupling, 
$\alpha_B$ and the unification scale $M_B$. At this stage we simply have two pairs of lines with different slopes 
intersecting at separate points. However, from the original six equations there are two remaining equations that can 
be used to predict the value of $\alpha_{A_3}$ and $\alpha_{B_3}$ at the higgsing scale. Remarkably we will see 
that these predictions agree with the measured value of $\alpha_3(M_Z)$ just like in the MSSM! This prediction 
crucially relies on the $\beta$ function coefficients, which are determined from the matter content at each site.

At the $A$ site we have the MSSM gauge and Higgs content, but there is only one matter generation 
(corresponding to the third generation). This implies that the $A$ gauge couplings will unify like in the usual 
MSSM. However, in addition to the MSSM fields, there are also link fields and flavor fields which are charged 
under $G_A$. The link fields come in complete SU(5) multiplets (a set of five ${\bf 5+{\bar 5}}$), and therefore 
do not affect the relative running of the gauge couplings. The flavor fields are responsible for generating the 
hierarchical structure in the CKM matrix between the first two generations and the third generation. As discussed 
above, the required flavor fields have the quantum numbers of a complete ${\bf 5}+{\bar {\bf 5} }$ multiplet under 
$G_A$. This means that the relative running of the $A$ gauge couplings is again not affected. Therefore the total 
contribution to the $\beta$ function coefficient at the messenger scale is $b_{A_i}= (43/5, 3,-1)$ which gives rise to 
$R_A=R=5/7$. The relative running is just like in the usual MSSM and the prediction for $\alpha_{A_3}^{-1}$ at the 
scale $\mu_0=\langle \chi\rangle$ is then given by an equation for the $A$ couplings similar to (\ref{a3inv}).

On the $B$ site, we still have the MSSM gauge multiplets and two generations of matter chiral multiplets, 
but there are no Higgs multiplets. Thus it would seem that the relative running of the $B$ couplings differs
from that in the MSSM. However, the $B$ site contains additional fields charged under $G_B$ which
are not in complete SU(5) multiplets. In particular, to generate higher-dimension terms in the superpotential 
responsible for the Yukawa textures, the $B$ site crucially contains a vector-like pair of $SU(2)$ doublets 
discussed earlier. These have precisely the quantum numbers of the MSSM Higgs doublets, and therefore 
with this additional contribution, the relative running of the $g_{B_i}$ couplings is exactly the same as in 
the MSSM! Moreover the link fields contribute a set of five ${\bf 5+{\bar 5}}$ multiplets, and there is also at 
least one pair of ${\bf 5+{\bar 5}}$ messenger fields responsible for gauge mediation of the supersymmetry 
breaking.  The total contribution to the $\beta$-function coefficients at the messenger scale is 
$b_{B_i}= (53/5, 5, 1)$. This gives rise to $R_B=R=5/7$, the same value as in the MSSM. The prediction 
for $\alpha_{B_3}^{-1}$ at the scale $\mu_0=\langle \chi\rangle$ is then given by an equation for the $B$ 
couplings similar to (\ref{a3inv}).

The prediction for $\alpha_3^{-1}$ at the higgsing scale then follows from combining the $A$ and $B$
values using (\ref{eqn:couplings}). Since both  $\alpha_{A_3}^{-1}$ and $\alpha_{B_3}^{-1}$ depend on the
same MSSM value of $R$ this leads precisely to the prediction (\ref{a3inv}), as in the MSSM!  Thus we have 
achieved a ``double" unification of both the $G_A$ and $G_B$ gauge groups, where the apparent unification of the 
low-energy gauge couplings splits into two sets of gauge couplings which separately unify. This is qualitatively 
similar to a two-site version of accelerated unification~\cite{ArkaniHamed:2001vr}, where the matter content at 
each site is determined by the flavor structure. Our two-site unified model therefore provides a nontrivial connection 
between unification and the flavor structure of natural supersymmetry.

\begin{figure}[t]
   \centering
   \includegraphics[width=5in]{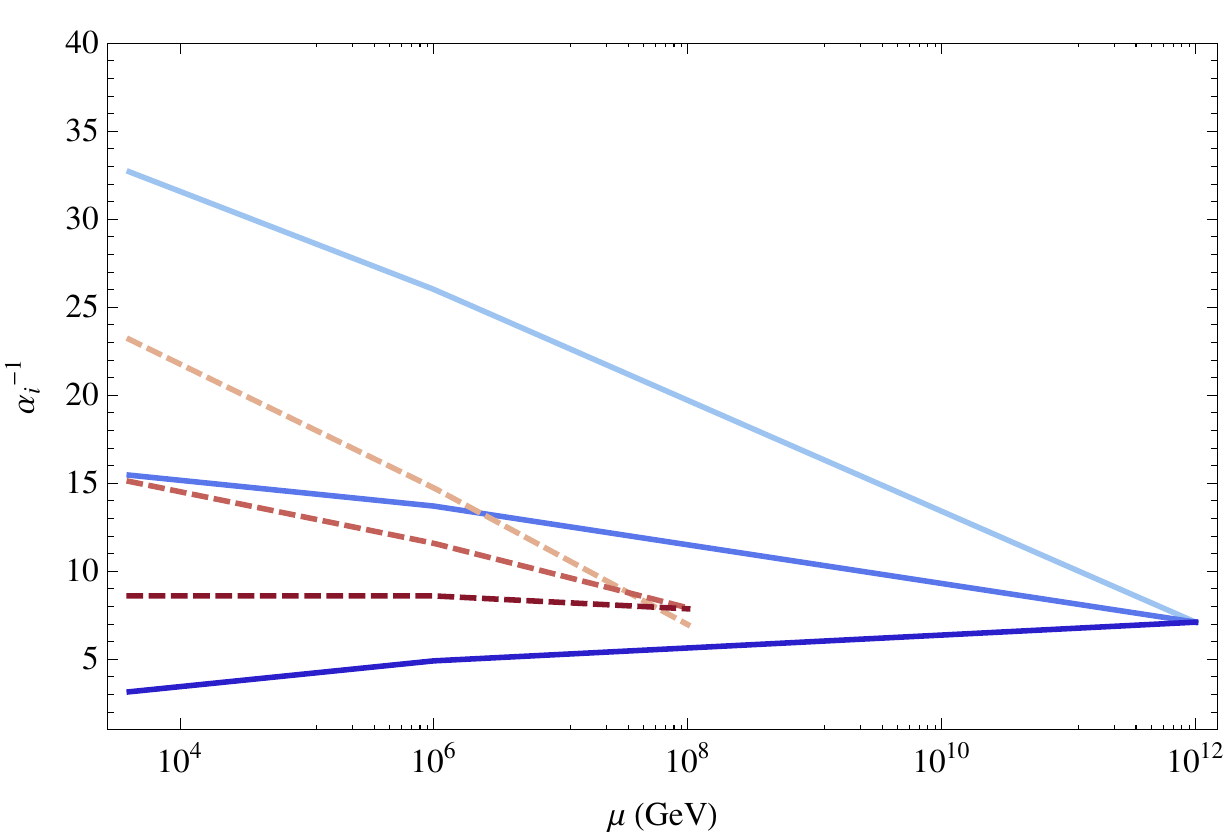}
   \caption{The unification of the $G_A$ (solid) and $G_B$ (dashed) gauge couplings. The higgsing scale is assumed to be
   4 TeV and the messenger scale is 1000 TeV. Here the $G_A$ couplings are taken to unify exactly at $M_A = 10^{12}$ GeV, while the $G_B$ couplings are run up from the higgsing scale.}
   \label{fig:gcu}
\end{figure}

The two unification scales $M_A, M_B$ for the $G_A, G_B$ couplings, respectively are related to the
higgsing scale and $M_{GUT}$ via the relation
\beq
\label{unifrelation}
        M_A M_B = \langle \chi \rangle M_{GUT}~.
\eeq
If the higgsing scale $\langle \chi \rangle = 10$ TeV we see from (\ref{unifrelation}) that
at least one of the unification scales is always lower than the usual $M_{GUT}$ value. 
This can lead to proton decay problems from dimension 5 and 6 operators that need to be 
addressed.

\subsubsection{Physics at the GUT scale(s)}

Ultimately, low-energy measurements point to three predictions for physics at high scales: the Standard Model gauge couplings  unify at a high scale; the Yukawa couplings of the bottom quark and $\tau$ lepton (but not the quark and lepton Yukawas of lighter generations) appear to unify at the same scale; and contributions to proton decay from unification-scale physics are small. These predictions imply that the theory at high energies takes the form of a four-dimensional deconstructed orbifold structure with UV gauge group $\left[(SU(3)\times SU(2)\times U(1)) \times SU(5) \right]^2$. Just as enlarging the MSSM gauge group $SU(3) \times SU(2) \times U(1)$ to $(SU(3)\times SU(2)\times U(1)) \times SU(5)$ at the GUT scale resolves all of the problems of simple $SU(5)$ GUTs while preserving gauge coupling unification, in this case a simple generalization preserves the prediction of double unification while keeping the proton stable.  Above the unification scales $M_A, M_B$ $G_A \times G_B \equiv [SU(3) \times SU(2) \times U(1)]^2$ is enlarged into $(G_{A_1} \times G_{A_2}) \times (G_{B_1} \times G_{B_2}) \equiv \left[(SU(3)\times SU(2)\times U(1)) \times SU(5) \right]^2$, with a set of bifundamental link fields charged under $G_{A_1} \times G_{A_2} \,(G_{B_1} \times G_{B_2})$. At the scale $M_A$, the first set of link fields break $G_{A_1} \times G_{A_2} \to G_A$, and likewise at the scale $M_B$ the second set break $G_{B_1} \times G_{B_2} \to G_B$.  If the gauge couplings of $G_{A_1}$ are larger than those of $G_{A_2}$ (and similarly for $G_{B_1}$ and $G_{B_2}$), then matching at the scales $M_A, M_B$  preserves the double unification prediction for $G_A$ and $G_B$ apparent from the infrared; the corrections from the non-unified $G_{A_1}, G_{B_1}$ couplings may be as small as a few percent. Standard Model matter fields charged under $G_A$ ($G_B$) at intermediate scales may be charged under either of the $G_{A_i}$ ($G_{B_i}$) at the unification scale depending on whether they appear to transform in unified multiplets.

Low energy data suggests $M_A \gtrsim 10^{12}$ GeV (and hence, following (\ref{unifrelation}), $M_B \lesssim 10^{8}$ GeV).
In particular, the successful prediction of $b - \tau$ unification implies the 3rd generation matter fields unify into complete $SU(5)$ multiplets, and so are charged under $G_{A_2} \equiv SU(5)$ at high energies. Proton stability forbids the Higgses from doing the same, suggesting they are charged under $G_{A_1} \equiv SU(3) \times SU(2) \times U(1)$, providing a natural explanation for doublet-triplet splitting and the absence of dimension-5 proton decay. Although the $X_A,Y_A$ gauge bosons couple 
only to the third generation fermions, proton decay can occur via CKM mixing to the first two generations. The scale of 
$G_A$ unification can therefore be as low as $\sim 10^{12}$ GeV while remaining consistent with limits 
on the rate of proton decay due to CKM suppression of the relevant processes. On the $B$ side, both dimension-5 and dimension-6 contributions to proton decay are addressed if the first two generations and the incomplete multiplets required by flavor are charged under $G_{B_1} \equiv SU(3) \times SU(2) \times U(1)$. Although this UV completion involves separate unification scales, it preserves the three infrared hints of grand unification: proton stability and the apparent unification both of gauge couplings and of $b, \tau$ Yukawa couplings in the ultraviolet, with no unsuccessful predictions for Yukawa unification of the first and second generations.

\subsubsection{Unification variations}

Of course, there are a variety of other possibilities for unification-scale physics. A simple alternative that preserves proton stability and gauge coupling unification with a single scale involves a 4D orbifold structure arising at a common unification value $M_A=M_B\equiv M_U\sim 10^{10}$ GeV. In this case, all the matter fields charged under $G_A (G_B)$ at low energies are charged under the non-unified $G_{A_1}(G_{B_1})$ in the UV, and there is a set of bifundamental link fields charged under $G_{A_1} \times G_{A_2} \,(G_{B_1} \times G_{B_2})$. At the scale $M_U$, these link fields break $G_{A_1} \times G_{A_2} \to G_A$ and $G_{B_1} \times G_{B_2} \to G_B$.  The choice of UV charge assignments explains the absence of prohibitive dimension-6 proton decay from $X, Y$ gauge bosons. The only relevant gauge bosons in this case are $X_{A_2,B_2}, Y_{A_2,B_2}$, which have no tree-level couplings to the Standard Model matter fields. Likewise, there is no need for triplet partners of the Higgs doublets, solving the doublet-triplet splitting problem and eliminating a possible source of dimension-5 proton decay. Indeed, given these interactions one may define an appropriate $U(1)$ baryon symmetry that separately commutes with the relevant generators of $G_{A_1} \times G_{A_2}$ and $G_{B_1} \times G_{B_2},$ protecting against contributions to proton decay from both gauge group sectors to all orders. Thus in this UV completion, unification occurs at an intermediate scale $\sim 10^{10}$ GeV, with unified gauge group $\left[(SU(3)\times SU(2)\times U(1)) \times SU(5) \right]^2$ where the matter fields are not charged under the $SU(5)$ groups.

Given a common unification scale $M_U$, it is also possible that an extra dimension opens up near the unification scale and the same attractive features are realized in a 5D orbifold GUT. Here somewhat more care is required due to potential contributions to proton decay from derivative-suppressed couplings to $X$ and $Y$ gauge bosons (absent in the two-site deconstruction), though in this case the rate is consistent with current bounds and may even be forbidden entirely by imposing an appropriate baryon symmetry.

There are also instances when the $G_B$ couplings hit a Landau pole at a scale $\Lambda_B < M_B$. In this case
an alternate --  albeit more speculative -- possibility is that the theory is UV completed by an asymptotically free dual description at high energies. Unification still occurs at the scale $M_B$ in the dual description, consistent with the IR prediction. In this case, the putative dimension-6 operators that induce proton decay in the IR map on to higher-dimensional baryonic operators in the UV theory, and dimension-6 proton decay is therefore suppressed by high powers of $\Lambda_B / M_B$ relative to the weakly-coupled expectation \cite{Abel:2008tx, Abel:2009bj}. Eventually there is even the amusing possibility that if $M_A > M_{GUT}$, 
then $M_B < \langle\chi\rangle$, in which case the $B$ couplings, which are only measurable above the higgsing 
scale, never actually unify. They in fact hit Landau poles, which may then unify in a dual perturbative UV description.

Finally, we note that gauge coupling unification on both sites frequently occurs before any couplings reach a Landau pole, despite the additional (unified) matter content associated with the two-site model. Even in the cases where a Landau pole is reached before the unification scale -- as is sometimes the case for $G_B$ if $M_A \sim M_B$ --  the ``strongification'' of $G_B$ does not affect the differential running of the $G_A$ couplings. Threshold corrections are also expected to be small,  as evidenced by the minuscule effect of QCD confinement on the running of $\alpha_{EM}$.

\section{The Higgs}

Any theory of natural supersymmetry in which the stop is light  must explain why the lightest CP-even neutral Higgs scalar lies above the LEP bound (and perhaps higher, near $\sim 125$ GeV). The unified two-site theory provides a natural explanation for the Higgs mass in the form of certain $D$-term corrections to the electroweak potential arising in the infrared. While in a generic deconstruction there is no reason for the $D$-term corrections to be significant, in this case the unification of both $G_A$ and $G_B$ points to $\mathcal{O}(1)$ corrections to the potential, forming an intimate connection between gauge coupling unification and the Higgs mass.

\subsection{$\mu$ and $B \mu$}

Any supersymmetric theory must provide an origin and scale for the $\mu$ term, $W \supset \mu H_u H_d,$ in order to explain electroweak symmetry breaking. Here the $\mu$ term has a particularly natural origin: the symmetries of the theory allow the irrelevant operator	
\beq\label{mu:C}
W \sim \frac{\chi \tilde \chi H_u H_d}{M_*} = \epsilon \langle \chi \rangle H_u H_d~. 
\eeq
which may be generated by the same physics that generates the fermionic flavor hierarchy. 
This produces a $\mu$ term of the proper order if $\langle \chi \rangle \sim 1-20$ TeV (since $\epsilon \sim 0.01 - 0.1$ from flavor considerations). Such a $\mu$ term is supersymmetric, so that the corresponding $B \mu$ term is naturally small; this theory has no $B \mu$ problem, although soft masses are generated by gauge/gaugino mediation.
For such an operator to explain the origin of the $\mu$ term, we must forbid the usual supersymmetric term $\mu H_u H_d$; this may be readily accomplished with a discrete symmetry. Since $B \mu$ is protected by both an $R$-symmetry and a $PQ$ symmetry, it is only generated radiatively below the scale of higgsing. The typical value of $B \mu$ is 
\beq
\label{eqn:Bmu}
B \mu \sim - \mu\left( \frac{3  \alpha_2}{2 \pi} m_{2} \log \frac{M_2}{m_{2}} + \frac{3}{5} \frac{\alpha_1}{2 \pi} m_{1}\log \frac{M_1}{m_{1}} \right)~,
\eeq
where $m_i$ is the gaugino mass of the $i$th Standard Model gauge group.

\subsection{The $D$-term}

A low scale of higgsing also solves one of the long-standing challenges of supersymmetric theories by raising the tree-level prediction for the lightest neutral Higgs mass. This effect arises upon integrating out the massive sgoldstone modes (the real scalar partners of the eaten goldstones), corresponding to the electroweak components of $\frac{1}{\sqrt{2}} {\rm Re} (\chi - \tilde \chi)$. In the supersymmetric limit this would simply yield the conventional MSSM $D$-terms, but in the presence of supersymmetry breaking it instead yields corrections to the $D$-terms of fields charged under $G_{A_i}$:
\beq
V =  \frac{g_{A_i}^2 g_{B_i}^2}{8(g_{A_i}^2 + g_{B_i}^2)} \left( 1 + \frac{g_{A_i}^2}{g_{B_i}^2} \frac{2 \tilde{m}_\chi^2 }{2(g_{A_i}^2 + g_{B_i}^2) \langle \chi \rangle^2 + 2\tilde{m}_\chi^2} \right) \left | \sum_j \Phi_{A_i}^{j \dag} T^i \Phi_{A_i}^j \right |^2~. 
\eeq   
where the $\{ \Phi_{A_i}^j \}$ are the scalar components of all fields charged under $G_{A_i}$. In particular, this leads to a sizable correction to the Higgs $D$-terms of the form
\beq \label{eqn:higgsD}
V_D = \frac{g^2}{8} \left( 1 + \Delta \right) \left |H_u^\dag \sigma^a H_u - H_d^\dag \sigma^a H_d \right |^2 + \frac{g'^2}{8} \left( 1 + \Delta' \right) \left |H_u^\dag H_u - H_d^\dag  H_d \right |^2
\eeq
where
\begin{equation}
\Delta =  \frac{g_{A_2}^2}{g_{B_2}^2} \frac{2 \tilde{m}_\chi^2}{M_2^2 + 2 \tilde{m}_\chi^2}  \hspace{1cm}
\Delta' =  \frac{g_{A_1}^2}{g_{B_1}^2}  \frac{2 \tilde{m}_\chi^2}{M_1^2 + 2 \tilde{m}_\chi^2}   \, .
\end{equation}
Here the unprimed quantities are those of the $SU(2)_A \times SU(2)_B \to SU(2)_L$, while primed quantities are those of $U(1)_A \times U(1)_B \to U(1)_Y$. 

{\it A priori}, there is no reason for these corrections to be large. However, the fact that both $G_A$ and $G_B$ unify perturbatively at a high scale and have similar beta function coefficients implies $g_{A_i} \simeq g_{B_i}$, so that the gauge coupling ratio appearing in $\Delta$ is naturally $\mathcal{O}(1)$. In this sense unification favors considerable corrections to the quartic, and hence the Higgs mass.

\subsection{Higgs potential}

The $D$-term correction to the Higgs potential modifies the tree-level Higgs spectrum, mixing angles, and couplings. The tree-level scalar potential for the Higgs doublets is given by the usual MSSM expressions with $g^2 \to g^2 (1 + \Delta)$ and $g'^2 \to g'^2 (1 + \Delta')$. Note that the usual stability and EWSB conditions of the MSSM are independent of the $D$-terms, so we still have the conventional constraints imposed by vacuum stability and a nonzero expectation value. It is straightforward to find the minimum of the Higgs scalar potential. The procedure is identical to that of the MSSM, except that there are additional $\mathcal{O}(\Delta)$ corrections from the $D$-term contribution to the scalar potential. To leading order we can neglect the corrections to the $Z$ mass, i.e., we can take $m_Z^2 \approx \frac{g^2+ g'^2}{2} v^2$, since the mixing corrections to $m_Z^2$ are much smaller than the $D$-term corrections by an amount $\approx \mathcal{O}(v^2/\tilde{m}_\chi^2) \ll 1$. At the minimum of the potential we can replace $B_\mu$ and $|\mu|$ with $\tan \beta \equiv v_u / v_d$. This leads to the usual expression for $\sin(2 \beta)$, as well as (in the limit of large $\tan \beta$) 
\beq \label{eq:ztune}
m_Z^2  + \frac{g^2 \Delta + g'^2 \Delta'}{2} v^2 = - 2 (m_{H_u}^2 + |\mu|^2) + \dots
\eeq
This makes clear one sense in which the little hierarchy problem is alleviated by the $D$-term correction by reducing the cancellation required between $m_{H_u}^2$ and $|\mu|^2$. We will discuss the naturalness of electroweak symmetry breaking in further detail below.

The tree-level prediction for the pseudoscalar mass $m_{A^0}^2$ is the same as in the MSSM, while the charged Higgs is shifted by an amount
\begin{eqnarray}
m_{H^\pm}^2 &=& m_{A^0}^2 + m_W^2 (1+ \Delta) ~.
\end{eqnarray}
Since limits on the pseudoscalar mass in the MSSM are obtained primarily from both direct and indirect limits on the charged Higgses, this correction allows for a lighter pseudoscalar than in the MSSM.

The mass eigenvalues for the real neutral Higgs scalars are related to those of the MSSM by taking $m_Z^2 \to m_Z^2 + \frac{g^2 \Delta + g'^2 \Delta'}{2} v^2$. Note this also shifts the mixing angle $\alpha$, which in this case is given by
\begin{eqnarray}
\frac{\tan 2 \alpha}{\tan 2 \beta} &=& \frac{m_{A^0}^2 + m_Z^2 +  \frac{g^2 \Delta + g'^2 \Delta'}{2} v^2}{m_{A^0}^2 - m_Z^2 -  \frac{g^2 \Delta + g'^2 \Delta'}{2} v^2}~.
\end{eqnarray}
This shift potentially alters the Higgs couplings relative to the MSSM. 

\subsection{Correction to the Higgs mass}
From the above, we see the tree-level prediction for the lightest neutral Higgs mass obtains a correction
\beq\label{eqn:mhtree}
(\delta m_h^2)_{\rm tree} \leq \frac{g^2 \Delta + g'^2 \Delta'}{2} v^2 \cos^2(2 \beta),
\eeq
saturated in the decoupling limit. The $\Delta$ corrections are $\mathcal{O}(1)$ when both $G_A$ and $G_B$ unify and the higgsing scale is low. As such, this constitutes a sizable contribution to the tree-level Higgs mass, allowing for a lightest neutral Higgs in the range $114-130$ GeV without requiring large logarithmic contributions from heavy or widely-split stops. Thus one fairly sharp prediction of this model is a Higgs boson in the range $114-130$ GeV with a light pair of stops below 1 TeV and small $A$-terms, since the stops are not required to lift the Higgs mass radiatively. 

\begin{figure}[t]
   \centering
   \includegraphics[width=4in]{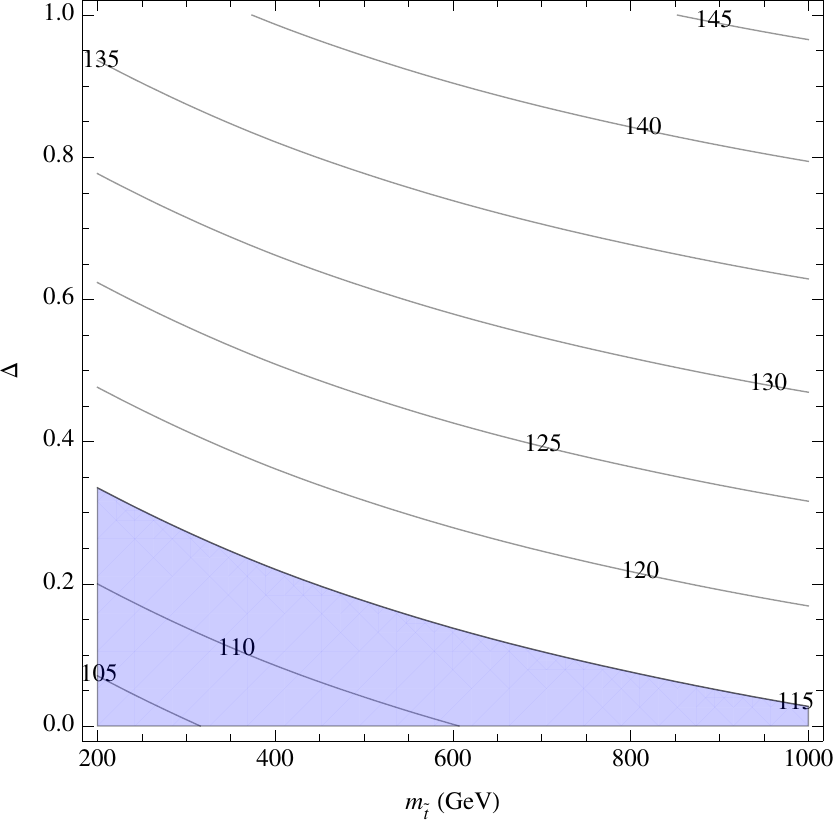}
   \caption{The lightest Higgs mass as a function of $m_{\tilde t}$ and $\Delta$ (for $\Delta' = \Delta$), including $D$-term and two-loop radiative corrections computed in {\tt FeynHiggs} \cite{Heinemeyer:1998yj}. The blue region is excluded by current LEP bounds. }
   \label{fig:mass}
\end{figure}

Typical values of the lightest neutral Higgs mass as a function of $m_{\tilde t} \,(\equiv \sqrt{m_{\tilde t_1}m_{\tilde t_2}})$ and $\Delta$ are shown in Fig.~\ref{fig:mass}. Here we have computed the tree-level mass with $D$-term contribution and the two-loop MSSM radiative correction using {\tt FeynHiggs} \cite{Heinemeyer:1998yj}; using only the one-loop radiative correction overestimates the Higgs mass by $\sim 5 - 10$ GeV. A Higgs mass near 125 GeV requires $D$-term corrections $\Delta \gtrsim 0.3$; such large corrections are favored by double gauge coupling unification and a low scale of higgsing.

\subsection{Naturalness}

There are {\it three} senses in which this theory improves the naturalness of electroweak symmetry breaking relative to the MSSM. In the MSSM, we characterize the tuning in electroweak symmetry breaking by the cancellation required between soft parameters to match the experimental value of the $Z$ mass. As mentioned above, the $Z$ mass relation is modified to (\ref{eq:ztune}) in the limit of large $\tan \beta$. Thus the additional positive-definite contribution on the LHS is the first sense in which naturalness is improved; a less precise cancellation between $m_{H_u}^2$ and $|\mu|^2$ is required.


The second improvement to naturalness comes from the increased prediction for the tree-level mass of the lightest neutral Higgs. In the MSSM, $m_h$ may be raised above the LEP bound only by relying on radiative corrections from the (necessarily large) stop mass. However, these same radiative corrections also increase the $H_u$ soft mass by an amount
\beq\label{eqn:radcorr}
\delta m_{H_u}^2 \sim - 12 \frac{y_t^2}{16 \pi^2} m_{\tilde t}^2 \log \frac{M_{UV}}{\mu_{IR}}
\eeq
where $M_{UV} \, (\mu_{IR})$ is the UV (IR) cutoff. This radiative correction increases the tuning required in relations such as (\ref{eq:ztune}). However, in our case the $D$-term correction raises the tree-level prediction for the Higgs mass as in (\ref{eqn:mhtree}), alleviating the need for a  heavy stop. Moreover, the split spectrum of scalars allows a light stop below 1 TeV to remain consistent with LHC limits.

The third improvement to naturalness comes from the lowered cutoff to radiative corrections. Whatever the size of the stop mass, the radiative correction (\ref{eqn:radcorr}) is sensitive to large values of $M_{UV}$. In the deconstruction, radiative corrections coming from the stop mass are cut off by the scale of higgsing, $\sim 10$ TeV, and are thus far smaller than in even the lowest-scale models of gauge mediation.

However, there is also a potentially new contribution to the degree of tuning coming from contributions of the link fields to the Higgs soft masses.  The link fields induce a loop correction to the Higgs soft masses of the form
\beq
\delta m_{H_u, H_d}^2 \simeq \left( \frac{3}{4} \frac{g_{A_2}^2}{g_{B_2}^2} \frac{g^2}{8 \pi^2} + \frac{1}{4} \frac{g_{A_1}^2}{g_{B_1}^2} \frac{g'^2}{8 \pi^2} \right) \tilde m_\chi^2 ~. 
\eeq
This is typically not problematic if the link fields are below $\sim 10$ TeV. Ultimately the various improvements to naturalness dominate, leaving the theory significantly less tuned than the MSSM for a fixed value of $m_{\tilde t}$ {\it even before considering the lightest Higgs mass}. Of course, the MSSM requires a heavy stop ($\gtrsim 1$ TeV) to accommodate the LEP bound on the Higgs mass, leading to an even further degree of fine-tuning.

\subsection{Higgs production and decays}

As the LHC nears the integrated luminosity required to discover or exclude the Standard Model Higgs over a wide range of masses, the question of Higgs production and decay rates becomes increasingly pressing. At low masses the most promising search channel at the LHC is the resonant diphoton channel, $gg \to h^0 \to \gamma \gamma$. As such, let us focus our attention on predictions for the relevant production and decay modes, gluon-gluon fusion ($gg \to h^0$) and the Higgs decay to two photons ($h^0 \to \gamma \gamma$).

The couplings of the lightest Higgs mass eigenstate $h^0$ to photons and gluons are not significantly altered for moderate to large values of $\tan \beta$. The $h^0 gg$ coupling arises almost entirely from top quark loops, while the $h^0 \gamma \gamma$ coupling arises from both top and vector boson loops. All $h^0 VV$ couplings are rescaled by $\sin^2(\alpha - \beta) \approx 1 + \mathcal{O}(\cot^2 \beta)$, while the $h^0 t \bar t$ coupling is rescaled by $\cos^2 \alpha / \sin^2 \beta \approx 1 + \mathcal{O}(\cot^2 \beta)$. Hence the effective $h^0 gg$ and $h^0 \gamma \gamma$ couplings are not significantly altered relative to the Standard Model prediction. The presence of stops running in the loop increases the $h^0 gg$ coupling by a factor of $\sim$ few percent, but is not significant if the stops are heavier than a few hundred GeV \cite{Arvanitaki:2011ck}.

Rather, the primary potential for change in the rate for $gg \to h^0 \to \gamma \gamma$ comes from the increase in the Higgs width due to enhanced coupling to down-type quarks and leptons. This enhancement is particularly significant for the couplings $h^0 b \bar b$ and $h^0 \tau^+ \tau^-$, which dominate the width at low Higgs masses. The branching ratios $BR(h^0 \to b \bar b, \, \tau^+ \tau^-)$ are enhanced by an amount $\sin^2 \alpha / \cos^2 \beta$ relative to the Standard Model. In the limit of large $\tan \beta$ and $m_A \gg m_Z$, this correction amounts to
\beq
BR(h^0 \to b \bar b) = BR(h^0 \to b \bar b)_{SM} \left[ 1 + \frac{4 m_Z^2}{m_{A^0}^2} + \frac{2 (g^2 \Delta + g'^2 \Delta') v^2}{m_{A^0}^2} + \dots \right]
\eeq
The first correction is the typical MSSM correction, which may be sizable if $m_A$ is not too large. This correction is further enhanced in our case by the $D$-term. Ultimately this is a sensitive test for mass splittings in the Higgs sector. Should the lightest Higgs emerge with a rate for $gg \to h^0 \to \gamma \gamma$ that is not too different from the Standard Model, this suggests that the pseudoscalar $A^0$, heavy neutral higgs $H^0$, and charged higgses $H^\pm$ are fairly heavy, above $\sim 400$ GeV. Conversely, deviations from the Standard Model inclusive rate may be indicative of lighter states in the extended Higgs sector. Note that here, as in the MSSM, it is difficult to increase the rate of exclusive diphoton processes such as vector boson fusion; we expect the VBF diphoton rate to be near the Standard Model value.

\section{Sparticles}

Apart from the signatures of the Higgs, the phenomenology of these theories is governed by a variety of states beyond the Standard Model. The most interesting are the stop, due to its role in maintaining the naturalness of electroweak symmetry breaking; the NLSP, which controls many of the collider signatures; and the variety of new states associated with the deconstruction that may lie within far LHC reach. Note that both the stop and the NLSP receive their soft masses from deconstructed gaugino mediation, in which the two-loop soft masses are suppressed by a factor $(\langle \chi \rangle / M)^2$ relative to the usual gauge-mediated contribution. However, there are also important three-loop contributions that are suppressed only by a loop factor, rather than  $(\langle \chi \rangle / M)^2$. These contributions dominate whenever $\langle \chi \rangle / M < 4 \pi$, as is the case here. Thus in what follows we will focus on the three-loop soft masses \cite{DeSimone:2008gm}, rather than the two-loop soft masses \cite{Auzzi:2010mb, Auzzi:2010xc}; the latter represent at most a few-percent correction.

\subsection{The Stop}

A principle preoccupation of natural theories of supersymmetry is the mass of the scalar partners of the top quark. These are by far the largest source of radiative corrections to both the Higgs soft masses and the physical mass of the lightest real Higgs scalar, and consequently the most sensitive measure of fine-tuning. 

In this theory the dominant soft masses for the left- and right-handed stop have been computed in \cite{DeSimone:2008gm} and depend on the gaugino, heavy gauge boson, and link field scalar soft masses. The soft masses of all fields charged under $G_A$ are of the same form, including for the Higgs doublets $H_u, H_d$. One might be concerned that the prospects for radiative electroweak symmetry are poor, given the lack of RG evolution over which to drive the $H_u$ soft mass negative.  However, the higher-loop contributions to the $H_u$ soft mass may still be considerable. In particular, there is a negative higher-loop contribution from stop and gluino loops to the Higgs soft mass. Due to the largeness of both the color charge and the top Yukawa coupling, these contributions may dominate over the leading-loop contribution and give rise to the negative soft mass required for electroweak symmetry breaking. Note that the higher-loop result is only significant for $H_u$; it requires both a large Yukawa coupling (in this case, $\lambda_t$) and no leading-loop color contribution.

\begin{figure}[h]
   \centering
   \includegraphics[width=4in]{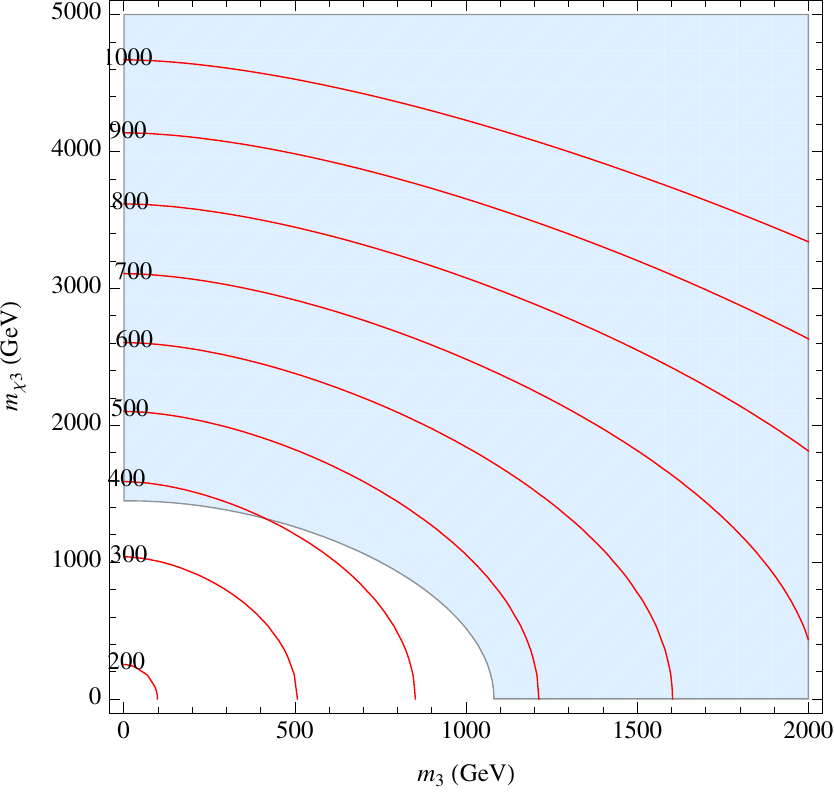}
   \caption{Region of viable EWSB relative to the stop mass as a function of gluino and colored scalar link field masses for a representative choice of parameters.   Red contours denote the mass of the lightest stop eigenstate $m_{\tilde t_1}$ in GeV, while the blue shaded area denotes $m_{H_u}^2 < 0$. Successful radiative electroweak symmetry breaking typically implies $m_{\tilde t_1} \gtrsim 400$ GeV. We have taken $m_{\chi_2} = 3$ TeV and $\langle \chi \rangle$ = 4 TeV, with the $G_A \times G_B$ gauge couplings corresponding to Figure 2. }
   \label{fig:stopmass}
\end{figure}

Intriguingly, this places an important indirect constraint on the mass scale of colored fields, particularly on the mass of the gluino and the colored link field scalars: the colored scales must be large enough that the higher-loop contribution to $H_u$ dominates over the leading-loop contribution. Since the mass of the stop is largely determined by these scales as well, the requirement of radiative electroweak symmetry breaking introduces a subtle constraint on the stop mass. Note that this is a general tension in any theory of natural supersymmetry breaking; although naturalness prefers the scale of colored fields to be low, successful electroweak symmetry breaking requires considerable radiative corrections. A representative example of the interplay between radiative electroweak symmetry breaking and the stop mass in this case is illustrated in Fig.~\ref{fig:stopmass}.

Thus these theories predict a naturally light stop, below a TeV, but not an {\it exceptionally} light stop; the natural range for the lightest stop mass is $400 \lesssim m_{\tilde t_1} \lesssim 1000$ GeV. This mass range is consistent with existing LHC limits at $\sim 2 $ fb$^{-1}$ provided $m_3 \gtrsim 700$ GeV \cite{Kats:2011qh, Brust:2011tb, Essig:2011qg, Papucci:2011wy}. 

\subsection{The NLSP}

The collider signatures  and dark matter candidates of supersymmetry are governed by the identity of the lightest supersymmetric particles. The lightest supersymmetric particle (LSP) in this scenario is invariably the gravitino, as supersymmetry breaking is mediated via gauge mediation to the $G_B$ sector. In low-scale scenarios, this corresponds to $F / M \sim 10^6$ GeV and hence $F \sim 10^{11} - 10^{12}$ GeV$^2$; the gravitino mass is thus in the eV - keV range, safe from cosmological difficulties. Structure formation forbids such light gravitinos from comprising the bulk of dark matter, as they freeze out when relativistic. Rather, to explain dark matter we require an additional, stable degree of freedom that is nonrelativistic at freezout. There are a variety of possible candidates for the cold dark matter particle in this case, including the axion and/or axino.

Although the gravitino lies at the end of every decay chain, the type of stable particles present in the final state is then governed by the {\it next-to-lightest} supersymmetric partner, the NLSP. Thus the identity of the NLSP is key to the dominant collider signatures of the theory. In this case the NLSP is typically the right-handed stau $(\tilde\tau_R)$, since it receives only a gaugino-mediated soft mass proportional to the hypercharge gauge coupling. However, depending on the value of $\mu$, it is also possible that the Higgsino could be the NLSP. This leads to various possibilities for collider searches.

A distinctive feature of the stau NLSP scenario is that the final state will be tau-rich. If the left-handed staus are also lighter than the 
neutralinos and charginos then decay chains will pass through the left-handed taus producing more taus in the final state. Since tau
decays produce soft leptons, limits remain weak from the Tevatron and the LHC. For example,  initial gluino production and decay 
via an intermediate wino state gives rise to the bound $m_{gluino}\gtrsim 700$ GeV~\cite{Kats:2011qh}. If instead the gluino decouples and the primary source of sparticles is electroweak production
then there are no limits on the wino-stau parameter space. In this case the stau NLSP may be observed in multilepton searches at the LHC \cite{CMS-PAS-SUS-11-013}. Alternatively for a Higgsino NLSP, two interesting cases were considered in Ref.~\cite{Kats:2011qh}, either Higgsino decay to a $Z$-boson plus gravitino or Higgs plus gravitino. Again, assuming gluino production,  the gluino mass lower bound is approximately 700 GeV, with no corresponding limit on the Higgsino mass.

\subsection{New states} \label{sec:pewk}

Provided the scale of higgsing is not too high, there are a number of additional light degrees of freedom with Standard Model quantum numbers that may be accessible at the LHC. This would provide a direct test of these low-scale models, in marked contrast to theories in which a large sfermion mass hierarchy is generated by unobservable dynamics at high scales.  In particular, the higgsing $G_A \times G_B \to SU(3)_c \times SU(2)_L \times U(1)_Y$ yields a complete set of  heavy $SU(3) \times SU(2) \times U(1)$ gauge bosons with masses $M_i^2 = 2 (g_{A_i}^2 + g_{B_i}^2 ) \langle \chi \rangle^2$. In principle the scale of these gauge bosons and their fermionic superpartners may be as low as a few TeV, putting them  within collider reach. Lower bounds on the gauge boson masses arise from precision electroweak and FCNC limits. 

How low may these new states lie? Existing constraints are dominated by limits from precision electroweak. In particular, the heavy $SU(2) \times U(1)$ gauge bosons and $SU(2)$ triplet scalars contribute to precision electroweak observables and the $\rho$ parameter. Their scale may be constrained by integrating out these fields and comparing the coefficients of the resulting irrelevant operators to experimental constraints. Limits are placed on various parameters by considering their contributions to Standard Model precision electroweak variables at the 95\% confidence level using the formalism of \cite{Han:2004az} with numerical data specialized to the case of a $U(2)$ flavor symmetry \cite{Han:2005pr}. By far the strongest constraint comes from the dimension-6 operator $|h^\dag D^\mu h|^2$, which constrains $(M_2^2 + 2 \tilde m_\chi^2) / g_{A_2}^2 \langle \chi \rangle \gtrsim 9900$ GeV; all other dimension-6 operators lead to considerably weaker limits \cite{njc}. Note that precision electroweak constraints in no way limit the scale of the heavy $SU(3)$ gauge boson. 

However, the heavy gauge bosons also contribute to flavor-changing neutral currents at tree level. In the case of the $SU(2) \times U(1)$ bosons the resulting limits are parametrically identical to, and weaker than, those arising from precision electroweak. However, the tree-level FCNCs induced by the heavy $SU(3)$ bosons provide new limits. Strictly speaking, the only inevitable contributions are MFV-like in nature, from which we may infer $M_3^2 / \langle \chi \rangle \gtrsim 1.9$ TeV at 95\% CL from $B_d - \overline B_d$ mixing. Consequently, the heavy $SU(3)$ gauge bosons may lie within reach at the LHC. However, in principle these theories predict a hierarchical right-handed mixing matrix as well as the left-handed CKM matrix, leading to contributions that are of NMFV form, for which more stringent bounds exist. In this case $M_3^2 / \langle \chi \rangle \gtrsim 9.6$ TeV, well beyond collider reach. Nonetheless, is plausible that these right-handed NMFV contributions are mitigated by additional textures or suppression in the right-handed quark mixings. Taken together, the limits coming from precision electroweak observables and flavor-changing neutral currents suggest that the various new states in this theory may lie low enough to be produced at the LHC, though they remain sufficiently heavy that their discovery will likely require significant integrated luminosity at 14 TeV.

\section{Conclusion}

Data is beginning to imperil natural theories of universal supersymmetry with MSSM-like electroweak symmetry breaking. Nonetheless, in view of precision electroweak measurements, a natural solution to the hierarchy problem suggests that supersymmetry should play some role in stabilizing the weak scale. In this paper we have presented a model that realizes the spectrum of natural supersymmetry by extending the Standard Model gauge group at 5-10 TeV. This leads to a $U(2)$ sflavor symmetry for the first two generations compatible with FCNCs, as well as an explanation for the broad features of Standard Model flavor. Gauge coupling unification is naturally preserved in the form of accelerated unification, with the appropriate differential running of gauge couplings guaranteed by the representations required by a successful theory of flavor. Although the unification scale(s) are lowered relative to that of the MSSM, the proton may remain stable while preserving the success of $b-\tau$ Yukawa coupling unification. The $\mu$ and $B \mu$ parameters of electroweak symmetry breaking have a natural origin, while the Higgs mass is explained by additional contributions to the quartic coupling, which are of the right size provided by perturbative gauge coupling unification.

\begin{table}[h]
\begin{center}
\begin{tabular}{|l|c|c|c|}
\hline
 & Deconstruction & Reconstruction & High scale \\ \hline
Naturalness & $\surd$ & $\surd$ & $\surd$ \\
Flavor & $\surd$ & $\surd$ & $-$ \\
Sflavor & $\surd$ & $-$ & $?$ \\
Higgs mass & $\surd$ & $-$ & $-$ \\
Unification & $\surd$  & ?  & ? \\ \hline
\end{tabular}
\end{center}
\caption{Comparison of theories of natural supersymmetry}
\label{virtues}
\end{table}%

It is useful to ask in what sense the model presented here differs from other theories of natural supersymmetry. We sketch a brief comparison in Table 1. Here ``reconstruction'' refers to theories of natural supersymmetry in a warped extra dimension \cite{Gherghetta:2003wm, Gabella:2007cp, Gherghetta:2011wc, Larsen:2012rq} and their  AdS/CFT ``dual'' versions \cite{ArkaniHamed:1997fq, Luty:1998vr,  Franco:2009wf, Craig:2009hf,  Sundrum:2009gv}, of which the model presented here may be considered a weakly-coupled ``deconstructed'' sibling. Yet there are differences; in ``reconstructed'' models, the $U(2)$ sflavor symmetry (which in four dimensions is assured by gauge mediation) is typically absent, as is a natural inherent correction to the Higgs quartic,\footnote{An interesting exception is \cite{Csaki:2012fh}, where the Higgs mass is also explained by compositeness.}
 while gauge coupling unification suffers from large threshold corrections or Landau poles.
Similarly, ``high scale'' refers to a broad collection of theories that generate the natural spectrum through RG effects or otherwise-decoupled dynamics (see, e.g., \cite{Dine:1993np,Pouliot:1993zm,Binetruy:1996uv,Dvali:1996rj,Delgado:2011kr}). Such theories invariably lack an explanation for the Higgs mass, and typically do not explain Standard Model flavor. They may lead to a $U(2)$ sflavor symmetry for the first two generations, though frequently this is achieved by arbitrarily distinguishing the first two generations for reasons unrelated to SM flavor. A noteworthy exception are theories of flavor mediation, in which the sflavor hierarchy is directly tied to SM flavor in such a way that preserves an accidental $U(2)$ symmetry \cite{Craig:2012yd}, though here too an explanation for the Higgs mass is absent.

\section*{Acknowledgments}

We thank 
John March-Russell for collaboration at early stages of this work. We also thank Nima Arkani-Hamed, Asimina Arvanitaki, Willy Fischler, Gian Giudice, Daniel Green, Andrey Katz, Patrick Meade, Alex Pomarol, and Giovanni Villadoro for useful discussions. This work was supported in part by ERC grant BSMOXFORD no. 228169. NC is supported in part by the NSF under grant PHY-0907744 and gratefully acknowledges support from the Institute for Advanced Study as well as hospitality from the Stanford Institute for Theoretical Physics, the Theory Division at CERN, and the Weinberg Theory Group at the University of Texas, Austin.  TG is supported in part by the Australian Research Council and gratefully acknowledges hospitality from the Stanford Institute for Theoretical Physics and the Theory Division at CERN.

\bibliographystyle{JHEP}
\bibliography{bib}

\end{document}